\documentclass[journal]{IEEEtran}
\usepackage{color}
\ifCLASSINFOpdf
   \usepackage[pdftex]{graphicx}
   \usepackage{array}
   \usepackage{multirow}
   \usepackage{hhline}
   \usepackage{bm}
   \usepackage{amsmath}

   \usepackage{float}
   \usepackage[ruled,vlined]{algorithm2e}
\usepackage{times}
\usepackage{cite}
\usepackage[para]{threeparttable}
\usepackage{array,booktabs,longtable,tabularx}
\newcolumntype{L}{>{\raggedright\arraybackslash}X}
\usepackage{ltablex}
\usepackage{siunitx}
\usepackage[flushleft]{threeparttablex}
\usepackage[dvipsnames]{xcolor}
\usepackage{xcolor, soul}
\sethlcolor{blue}

\usepackage{etoolbox}
\makeatletter
\patchcmd{\@maketitle}
  {\addvspace{0.5\baselineskip}\egroup}
  {\addvspace{-1\baselineskip}\egroup}
  {}
  {}
\makeatother

   \graphicspath{{../pdf/}{../jpeg/}}

   \DeclareGraphicsExtensions{.eps,.pdf,.jpeg,.png}
\else

\fi

\begin{document}
\title{AIDX: Adaptive Inference Scheme to Mitigate State-Drift in Memristive VMM Accelerators}
\author{Tony Liu, Amirali Amirsoleimani, Fabien Alibart, Serge Ecoffey, Dominique Drouin, and Roman Genov

\thanks{Tony Liu, Amirali Amirsoleimani, and Roman Genov are with the Department of
Electrical and Computer Engineering, University of Toronto, 10 King’s College Road, Toronto, Ontario, Canada. Fabien Alibart, Serge Ecoffey and Dominique Drouin are with the Interdisciplinary Institute for Technological Innovation - 3IT, University of Sherbrooke, Qc, Canada. (A. Amirsoleimani corresponding author email: amirali.amirsoleimani@utoronto.ca).}}

\markboth{Submitted to IEEE TRANSACTIONS ON CIRCUITS AND SYSTEMS-II: EXPRESS BRIEFS}
{Shell \MakeLowercase{\textit{et al.}}: Bare Demo of IEEEtran.cls for IEEE Journals}

\maketitle

\begin{abstract}
An adaptive inference method for crossbar (AIDX) is presented based on an optimization scheme for adjusting the duration and amplitude of input voltage pulses. AIDX minimizes the long-term effects of memristance drift on artificial neural network accuracy. The sub-threshold behavior of memristor has been modeled and verified by comparing with fabricated device data. The proposed method has been evaluated by testing on different network structures and applications, e.g., image reconstruction and classification tasks. The results showed an average of $60\%$ improvement in convolutional neural network (CNN) performance on CIFAR10 after $10000$ inference operations as well as $78.6\%$ error reduction in image reconstruction.
\end{abstract}
\vspace{-0.1cm}
\vspace{-0.3cm}
\IEEEpeerreviewmaketitle

\section{Introduction}

\IEEEPARstart{R}{resistive} switching memory crossbars have emerged as potentially high-speed and low-power accelerators for vector-matrix multiplication (VMM) \cite{0,0a}. However, non-idealities and defects in these platforms dramatically impact the neural network (NN) performance and accuracy. One of the significant and not extensively studied non-ideal phenomena is memristance drift~\cite{1} and it occurs in different types of resistive switching memory technologies in various ways. For instance, phase change memories (PCM) will experience increasing resistance due to drift, even when there is no voltage applied over the cell \cite{2a}. On the other hand, for memristors, state-drift from their programmed state happens as a result of many repeated VMM operations which leads to the computational accuracy degradation (Fig. 1). 
\begin{figure*}[!t]
    \centering
    \includegraphics[width=\linewidth]{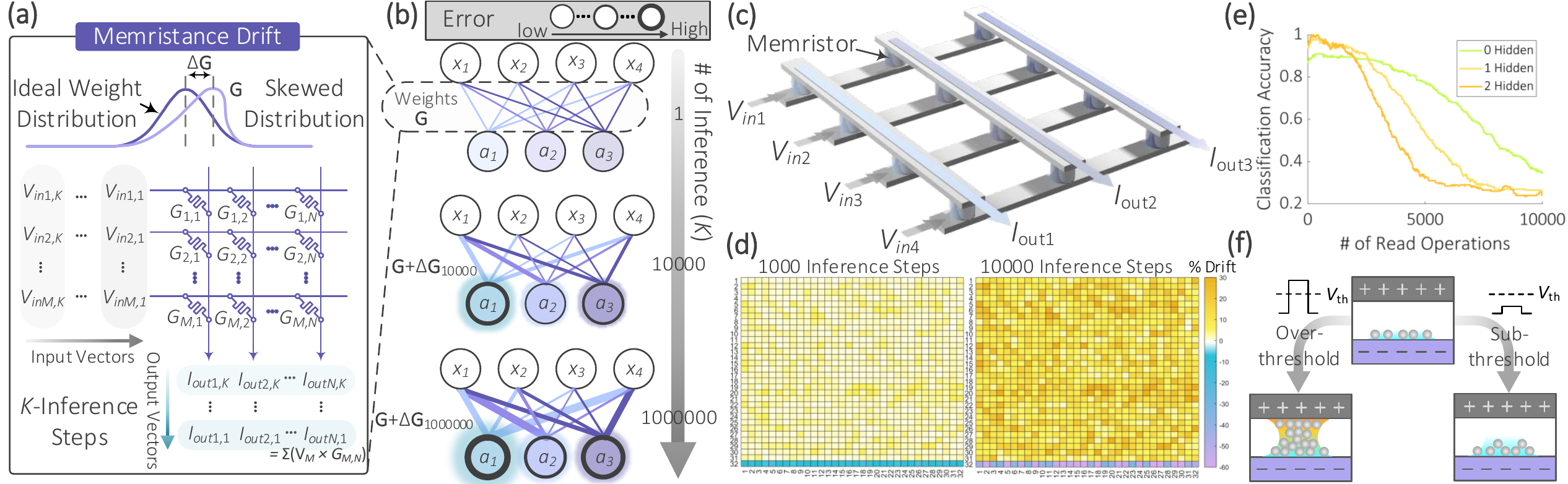}
    \vspace{-0.7cm}
    \caption{(a) Neural network (NN) forward pass VMM performed on a memristor crossbar. (b) Degradation of neural network weights and output accuracy over time. (c) 3D memristor crossbar array structure. (d) Percentage change of memristors conductance in a randomly chosen subset of NN weights done on a simulated memristor crossbar. (e) Impact of simulated state-drift on MNIST classification accuracy with varying number of hidden layer. (f) Sub- and over-threshold behaviour of device transition oxide filament.}
     \vspace{-0.4cm}
    \label{fig:Post Simulation Results}
\end{figure*}
Previous studies \cite{2,3,4} on memristance drift in memristor technology have mainly been focused on high-density memory where memristors are used solely for storage rather than computation. 
More recent reports on drift\cite{5a} for computational memristor crossbars include an inline calibration approach \cite{5} which involves optimizing the calibration time of the memristor crossbar. By performing polynomial fitting on the computational error data, a $21.77\%$ calibration efficiency is achieved.
A closed-loop weight compensation based solution is presented in \cite{6} which minimizes the effects of state-drift by increasing the computational service lifetime by $14.95\times$  and results in approximately $70\%$ computational accuracy degradation within $1705$ read operations. In this brief, we present an adaptive inference scheme (AIDX) as a flexible optimization procedure that automatically adapts to existing crossbar non-idealities and circuit parasitics and can be applied to any VMM-based task. According to experimentally verified simulations, AIDX cuts accuracy loss due to the device state-drift in modern convolutional neural networks (CNN) by more than $60\%$ as well a $78.6\%$ error reduction in image reconstruction.
\vspace{-0.3cm}
\section{Preliminary}
\subsection{Impact of Memristance Drift on Crossbar MAC Operations}
Memristance drift is defined as the unintended small changes in memristor conductance caused by a low-voltage read/inference operation. 
Ideally, for the ideal weight distribution $(\textbf{G})$ the output current $j$-th column $I_j$ is given by $I_j = \sum_{i} G_{ij} V_{i} $ (Fig. 1(a)). We can define the memristance drift caused by the $k$-th inference operation as $\delta G_k$ and the conductance of the memristor at the $(k+1)$-th iteration as:
\begin{equation}
    G_{k+1} = G_{k} + \delta G_{k}
\end{equation}
The total memristance drift due to the $k$-th operation is $\Delta G = \sum_{i}^{k} \delta G_{i}$. As such, the real output current of the $j$-th column at the $k$-th operation is $I'_j = \sum_i (G_{ij} + \Delta G_{ij}) V_i$. The current error $I - I'$ due to memristance drift can be quite problematic in larger crossbars because current scales with crossbar size. However, a differential mapping scheme can prevent the build-up of memristance drift error in very large arrays because the error in the positive column will scale at the same rate as the negative column. Fig. 1(b) illustrates the concept of small changes in NN weights accumulating into much larger errors in the output layer. Fig. 1(c) illustrates 3D structure of the network in Fig. 1(b). Fig. 1(d) shows sample heatmaps of simulated $32\times32$ array of memristors conductance changes due to memristance drift.The bottom row represents the bias weights of a NN and they are initially mapped to a high-conductance state which is why it is the only row with reduction in overall conductance. Fig. 1(e) examines the state-drift impact on MNIST classification task for multi layer perceptron network with various number of hidden layers and Fig. 1(f) illustrates the difference between above- and sub-threshold memristor switching. 
\vspace{-0.3cm}
\subsection{Memristance drift modeling and analysis}

Behaviour-based memristor models are typically used in memristor crossbar simulations due to their simplicity and light computational load.  However, most behaviour-based models do not consider memristance drift and approximate the internal state change due to an applied sub-threshold voltage to be zero. To address this issue, we propose an extension to the popular VTEAM model \cite{8} that accounts for the minute changes in internal state due to sub-threshold voltages. 
For model consistency, we adopt a similar mathematical structure in the sub- and above-threshold region:
\begin{equation}
    \frac{dw(t)}{dt}= 
\begin{cases}
    k_{s,off} \cdot \left( \frac{v(t)}{v_{off}}\right)^{\alpha_{s,off}} \cdot f_{s,off}(w),& \text{if } 0 \leq v < v_{off}\\
    k_{s,on} \cdot \left(\frac{v(t)}{v_{on}}\right)^{\alpha_{s,on}} \cdot f_{s,on}(w),& \text{if } v_{on} < v < 0
\end{cases}
\end{equation}
Here, $v_{off}$ and $v_{on}$ represent the RESET and SET voltage thresholds respectively. $w(t)$ is the internal state variable and is related to the resistance $R$ of the memristor as $R(t) = R_{off} w(t) + R_{on} (1 - w(t))$. $k_{s,off}$ and $k_{s,on}$ are fitting parameters that represent the rate of ion migration at any given applied sub-threshold voltage. Similarly, $\alpha_{s,off}$ and $\alpha_{s,on}$ are parameters that characterize the exponential relationship between speed of ion migration and the applied voltage. $f_{s,on}(w)$ and $f_{s,off}(w)$ are window functions that bounds the state between $0$ and $1$.
The time derivative of the resistance can be expressed as:
\begin{equation}
   \frac{dR(t)}{dt} = R_{off} \frac{dw(t)}{dt} - R_{on} \frac{dw(t)}{dt}
\end{equation}
 $R_{on}$ and $R_{off}$ are low and high resistance state of device, respectively. Cycle-to-cycle and device-to-device variations in sub-threshold drift speed are modelled by adding 15\% random Gaussian noise to $k$ and $\alpha$ parameters. The probability density function (PDF) of $k_{on}$ is shown in Eqn. (4) where $k_{on}$ is the ideal, fitted parameter and $x$ represents $k_{on}$ with added Gaussian noise.  The PDF of the other $k$ and $\alpha$ parameters follow the same structure as Eqn. (4).
\begin{equation}
    f_{kon}(x) = \frac{1}{\sqrt{0.3k_{on}\pi}} e^{-\frac{(x-k_{on})^2}{0.3k_{on}}}
\end{equation}
To validate our proposed model, the VTEAM extension is applied to TiOx-based memristor device (Fig. 2(a)) data. The extended VTEAM $k$ and $\alpha$ parameters were fit using simulated annealing algorithms and gradient descent with SET and RESET voltage thresholds of $-0.6$V and $0.6$V. Fig. 2(b-c) illustrates that the extended VTEAM models sub-threshold memristor behaviour much more accurately than the original VTEAM. Fig. 2(d) shows a 3D plot of how memristor switching behavior and conductance changes with internal state $w$ and applied voltage in sub-threshold region.


\begin{figure}[!t]
    \centering
    \includegraphics[width=0.5\textwidth]{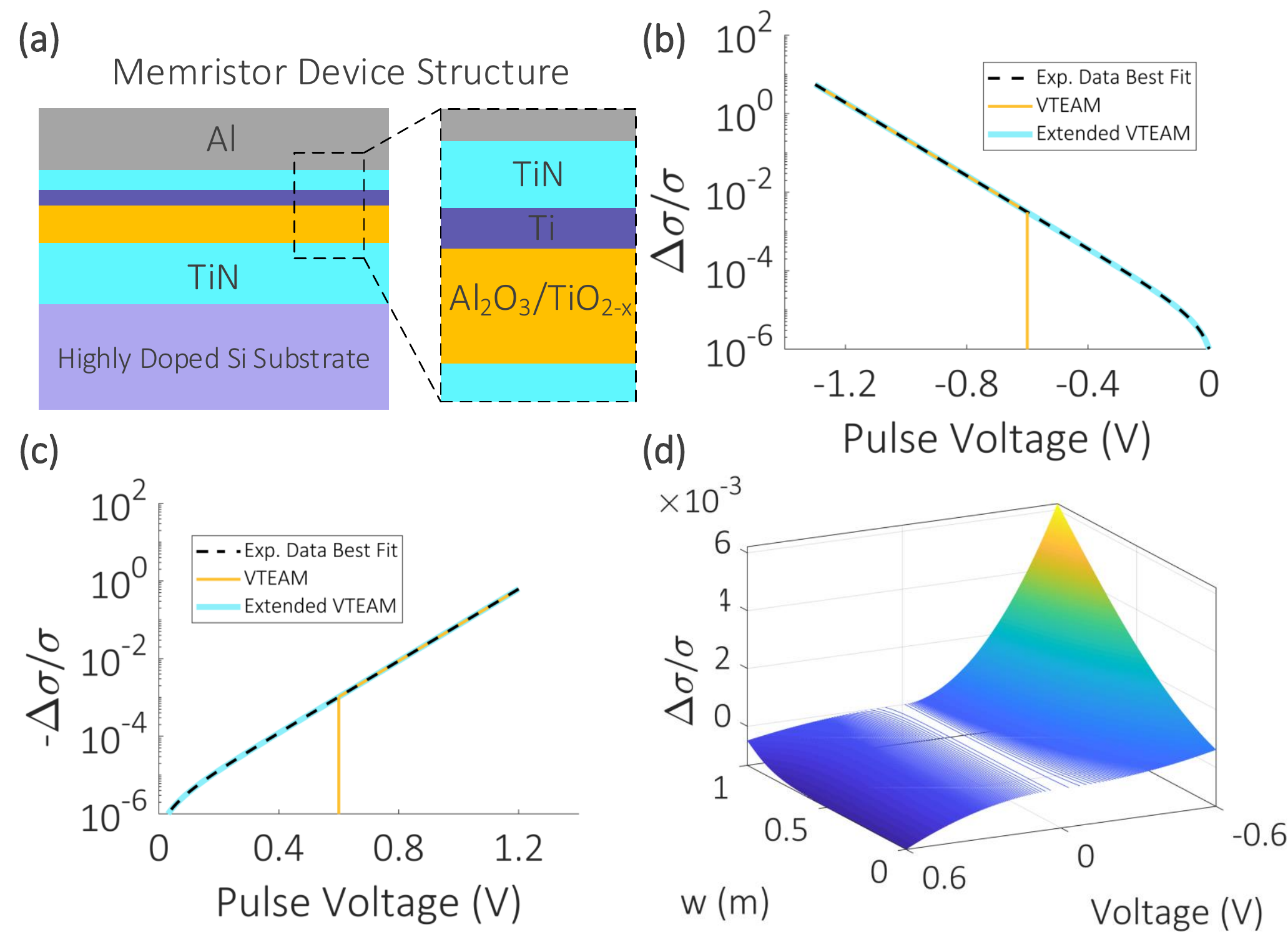}
    \vspace{-0.7cm}
    \caption{(a) Memristor device structure used for fitting. (b) Comparison of $I$-$V$ fitting of the extended VTEAM with experimental data and the original model in sub-threshold region for SET operation ($k_{s,on} = -8.445\times10^{-6}$, $\alpha_{s,on} = 6$). For clarity, we extracted a best fit curve to represent the experimental data from over the threshold and extrapolated the curve into sub-threshold region by keeping the gradual drift trend.} (c) Fitting comparison with experimental data for RESET operation ($k_{s,off} = 1.126\times10^{-7}$, $\alpha_{s,off} = 5$). (d) 3D characterization of the extended VTEAM for the same device with respect to the internal state variable $w$ and voltage.
    \vspace{-0.4cm}
    \label{fig:Baseline Bit Acc}
\end{figure}
\vspace{-0.2cm}
\section{Methodology}
\subsection{Problem and Assumptions}

By formulating the issue of memristance drift as an optimization problem, we can develop an optimization scheme to minimize accuracy degradation. With no memristance drift, the ideal mean squared error (MSE) is $E_0 = \sum_j (y_j - \sum_i G_{ij}V_i)^2$ and the real MSE at the $k$-th operation is $E_k = \sum_j (y_j - \sum_i (G_{ij} + \Delta G_{ij})V_i)^2$. Where $V_{i}$ is the voltage applied to $i$-th row and $\Delta G_{ij}$ is the total memristance drift of the $ij$-th memristor from its originally programmed value. We define the error due to memristance drift $E_{Drift}$ as the difference in MSE between the initially programmed state ($E_0$) and the $k$-th inference operation ($E_k$).
As an optimization problem, the goal is to minimize the increase of $E_{\mathrm{Drift}}$ with respect to time. 
The change in conductance due to memristance drift, $\Delta G$, can mainly be optimized by input to voltage amplitude mapping and relative inference voltage pulse width. Factors that cannot be easily changed such as specific memristor characteristics and overall crossbar structure will be ignored in the optimization procedure. 

\begin{figure*}[!t]
    \centering
    \includegraphics[width=1\textwidth]{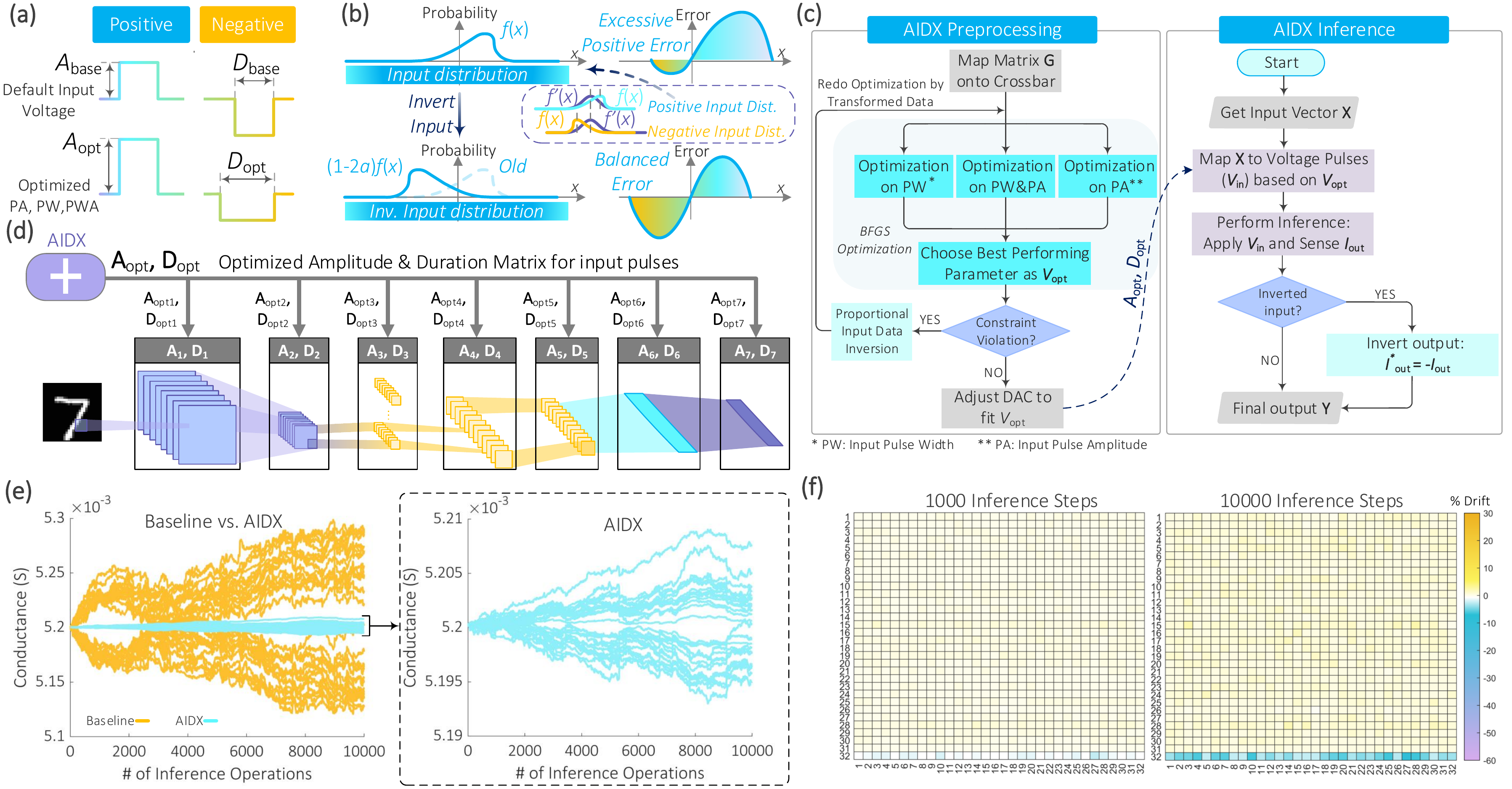}
    \vspace{-0.7cm}
    \caption{ (a) AIDX's optimized parameters \textbf{A} and \textbf{D} mapped onto input voltage pulses. (b) Proportional input inversion to balance memristance drift error and prevent constraint violations. (c) AIDX Design flowchart. (d) AIDX optimized parameters for duration ($D_{\mathrm{opt}}$) and amplitude ($A_{\mathrm{opt}}$) are applied to each CNN layer's input separately. (e) Simulated memristance drift with and without AIDX over time with 15\% device-to-device variations. To better observe the total state-drift, all devices’ initial conductance are set to $0.0052$ $\mathrm{S}$ and a positively skewed pre-generated random sequence of voltage pulses were applied to half the memristors and a negatively skewed pulses to the other half. (f) Percentage change in conductance of the same simulated memristors shown in Fig. 1(d) by utilizing AIDX.} 
    \label{fig:DesignFlow}
    \vspace{-0.5cm}
\end{figure*}
\vspace{-0.2cm}
\begin{algorithm}[!b]
\SetAlgoLined
 1. Obtain a direction $\mathbf{p}_{k}$  through solving $B_{k} \textbf{p}_{k}  = -\nabla f(\mathbf{x}_{k})$
 
 2. Perform $\mathrm{1D}$ Line Search to find step size $\alpha_{k}$ such that $\alpha_{k} = argmin f(x_{k} + \alpha \mathbf{p}_{k})$
 
 3. $\textbf{s}_{k} = \alpha_{k} \textbf{p}_{k}$
 
 4. $\mathbf{x}_{k+1} = \mathbf{x}_{k} + \mathbf{s}_{k}$
 
 5. $\textbf{y}_{k} = \nabla f(\mathbf{x}_{k+1}) - \nabla f(\mathbf{x}_{k})$
 
 6. $B_{k+1} = B_{k} + \frac{\mathbf{y}_{k} \mathbf{y}_{k}^{T}}{\mathbf{y}_{k}^{T} \mathbf{s}_{k}} - \frac {B_{k} \mathbf{s}_{k} \mathbf{s}_{k}^{T} B_{k}^{T}}{\mathbf{s}_{k}^{T} B_{k} \mathbf{s}_{k}}$ 
 
 7. Repeat 1-6 until $x$ converges.
 \caption{BFGS Algorithm}
\end{algorithm}

\vspace{-0.2cm}
\subsection{Optimization Methodology}

We will frame the minimization of $E_{Drift}$ as an unconstrained optimization problem where $\textbf{A}$ is the input to voltage amplitude mapping and $\textbf{D}$ is the relative voltage pulse width:
\begin{equation}
min_{A,D} E_{Drift} (\textbf{A},\textbf{D})
\end{equation}
$\textbf{D}$ is a vector that represents the length of a positive input read pulse relative to a negative read pulse. For instance, a given row can have a positive read pulse of $200 ns$ while the negative read pulse is only $150 ns$ long. Similarly, $\textbf{A}$ is a vector whose elements represent the relative inference voltage pulse amplitude ratio for positive to negative inputs (Fig. 3a).  In summary, AIDX modifies the amplitude and duration of inference voltage pulses to minimize memristance drift for a given task. Even the minimum allowable voltage pulse amplitude and widths will still result in noticeable memristance drift after many inference operations. As such, AIDX is required to minimize aggregate memristance drift through balancing the total drift in the SET and RESET directions. 
We use the popular Broyden-Fletcher-Goldfarb-Shannon (BFGS) algorithm \cite{9} for this optimization problem. The BFGS algorithm is a quasi-Newton method that relies on the gradient of the objective function to find the optimal solution. However, $E_{Drift}$ is an unknown function that can only be evaluated, so the $\nabla E_{Drift}$ had to be approximated using finite-difference approach. Gradient-free optimization methods like Nelder-Mead simplex method \cite{10} were also explored, but quasi-Newton algorithms were most effective for this problem.
\vspace{-0.2cm}
\subsection{Constraint Violations}
Normally, the optimized voltage amplitude ratio $\textbf{A}$ and width ratio $\textbf{D}$ can be reasonably used.  However, there are certain cases where elements of the optimal $\textbf{A}$ and $\textbf{D}$ are far too large or too small to be implemented practically, typically when the device characteristics and input distribution are heavily skewed. To address this issue, we will first frame the optimization problem through a different lens. Let's start with a simplified scenario of a single memristor with input data modelled by the discrete random variable $X$ with a probability density function (PDF) of $f(x)$. Defining the time derivative of the internal state $w$ for a given $x$ as 
\begin{equation}
    \frac{dw(x)}{dx} = g(x).
\end{equation}
In our sub-threshold model, $g(x)$ is the same as Eqn. (2) as $v(t)$ replaced with $v(x)$ which represents the mapping function of input $x$ to voltage amplitude $v(x)$. The average rate of memristance drift given input distribution $X$ is as follows:
\begin{equation}
E[\frac{dw(x)}{dx}] = \sum_{x} g(x) f(x).
\end{equation}
The optimization problem over $E_{Drift}$ can also be reframed as minimizing $\left|E[\frac{dw(x)}{dt}]\right|$ over all memristors where the $g(x)$ parameters for each memristor is sampled according to Eqn. (4) to account for device-to-device variations. The AIDX scheme defined so far only affects $g(x)$, but has not yet made any adjustments related to $f(x)$. If we allow AIDX to first optimize over $f(x)$, the issue of impractical $\textbf{A}$ or $\textbf{D}$ can be circumvented. One of the only useful recoverable transformations of the input vector $x$ is inversion through multiplying by $-1$. By inverting a random proportion $a$ of the input data, the input PDF is transformed into $f'(x)$:
\begin{equation}
    f'(x) = (1-2a)f(x), 0<a<1
\end{equation}
Impractical $\textbf{A}$ or $\textbf{D}$ only occur when either $\left|E[\frac{dw(x)}{dt}]\right| >> 0$ or $\left|E[\frac{dw(x)}{dt}]\right| << 0$ before applying AIDX.  As such, if we can optimize $\left|E[\frac{dw'(x,a)}{dt}]\right|$ over $a$, $\left|E[\frac{dw(x)}{dt}]\right|$ will be brought close to $0$ before optimizing $\textbf{A}$ and $\textbf{D}$ which will therefore prevent any constraint violations. Fig. 3(b) illustrates our approach to constraint violations. 
\subsection{General Solution Flow}
As it can be seen in Fig. 3(c), during pre-processing, optimization is done in three separate scenarios to guarantee optimal fitting parameters. Once hardware constraint violations are resolved with input data inversion, the input circuits e.g. digital-to-analog converters (DACs) are adjusted to fit the optimized input to voltage signal mapping parameters. The majority of AIDX takes place during pre-processing which only needs to be done once for any given task. The only difference in the AIDX inference operation as compared to a normal inference operation is to recover the intended output current from an inverted output through multiplying by $-1$.  Fig. 3(d) summarizes the general pipeline of AIDX. Fig. 3(e) shows the evolution of memristor state due to memristance drift for AIDX and the baseline model and Fig. 3(f) is a heatmap of a portion of the memristor crossbar at 1000 and 10000 inference steps where the bottom row represents the bias. While memristance drift is a phenomenon that can cause memristors to switch in both the set and reset direction as seen in Fig. 3(e), almost all memristors within a crossbar will typically drift in only one direction for image-based applications. These inputs are almost entirely positive which causes an aggregate drift in the reset direction. Other reasons for a unidirectional aggregate drift include: the device switching speed is not the same in the set and reset direction and most non-biased memristors are being initialized close to the high resistive state where the drift speed is strongly skewed in the set direction (Fig 2(d)).
\vspace{-0.2cm}
\section{Results and Discussion}
In this paper, all simulations are performed using our extended VTEAM memristor model \cite{11} by including the effects of sub-threshold state-drift whose parameters are fit according to the experimental data shown in Fig. 2. 
We integrate this memristor model into our existing 1T1R memristor crossbar simulation to simulate both memristance drift and crossbar non-idealities like sneak paths and line resistance. 
A differential weight mapping scheme is used where each element is mapped onto a pair of memristors where one memristor represents positive values and the other represents negative values. 
 In Fig. 4(a), AIDX is tested across 10 baseline tasks from the Proben1 benchmark datasets \cite{11}. We trained a shallow 1-hidden layer NN for all of these tasks. 
While there are large variations in baseline performance across different tasks, it should be noted that all baseline tasks ended at around the same classification accuracy as random guessing due to some tasks having more classification categories than others. 
\begin{figure*}[!t]
    \centering
    \includegraphics[width= \linewidth]{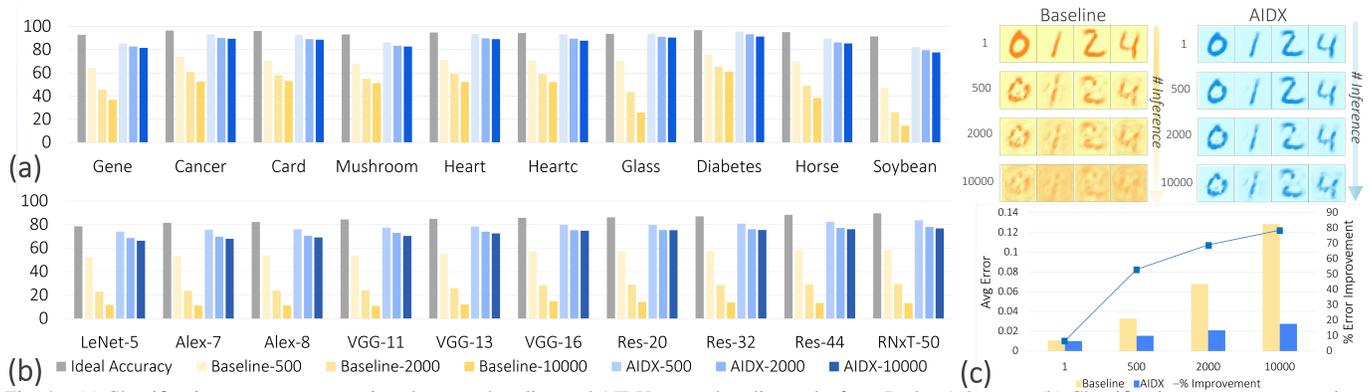}
    \vspace{-0.9cm}
    \caption{ (a) Classification accuracy comparison between baseline and AIDX across baseline tasks from Proben1 datasets. (b) Classification accuracy comparison across different CNN architectures on CIFAR-10 image classification dataset. (c) Sample reconstructed MNIST images and average image reconstruction error from baseline and AIDX-enhanced auto encoders.}
    \label{fig:General Results}
    \vspace{-0.5cm}
\end{figure*}
To verify our solution's effectiveness for more practical applications, we adopted AIDX for a selection of CNN architectures on the CIFAR10 dataset. The CNN memristor crossbar mapping scheme used is similar to the one found in \cite{12}. Fig. 4(b) compares the performance of 10 different CNN architectures between AIDX and the baseline model.  
As compared to the shallow NNs used for the $\mathrm{Proben1}$ datasets, the CNNs had an overall higher speed of accuracy degradation. The worse performance of CNNs is to be expected because of error propagation from one layer to the next amplifying the effect of memristance drift. The error in column $j$ of the $l+1-th$ layer in a fully connected NN is:
\begin{equation}
    E_{j,l+1} = \sum_i^n V_{i,l+1}(\sigma (E_{i,l}) + \Delta G_{ij,l+1})
\end{equation}
Here, $n$ is the total length of the input vector and $\sigma$ is the activation function of the $l-th$ layer. Due to the large number of parameters in modern CNNs, BFGS optimization in AIDX is performed sequentially layer by layer to reduce optimization time. Applying AIDX to the selected CNNs provided consistent improvements in classification accuracy on CIFAR10. 
The consistent improvement in AIDX performance across varying sizes and designs of CNNs demonstrates the proposed method flexibility across different crossbar sizes and structure.
In addition to classification tasks, we wanted to demonstrate AIDX's effectiveness in a different type of memristor crossbar application.  Fig. 4(c) shows the results of image reconstruction with the MNIST dataset.  For this task, a 1-hidden layer auto-encoder with $32$ hidden units was trained off-chip which corresponds to a $24.5\times$ compression factor. With AIDX, the average mean squared error has improved by $78.6\%$ over the baseline after $10000$ inference operations.
\vspace{-0.2cm}
\section{Overhead Analysis and Comparison}
Different state-drift mitigation techniques have been compared with two different AIDX configurations optimized for accuracy (AIDX-A) and power efficiency (AIDX-P) to implement a MLP network in Table 1. AIDX-A is the baseline AIDX method discussed in previous sections while AIDX-P adds a L2 regularization terms for $\textbf{A}$ and $\textbf{D}$ as follows: $min_{A,D} (E_{\mathrm{Drift}} (\textbf{A},\textbf{D}) + \lambda_1 \sum A^2 + \lambda_2 \sum D^2)$. Where $\lambda_1$ and $\lambda_2$ are regularization constants and regularizing the voltage amplitude and width ratios allows AIDX-P to reduce the passive crossbar power consumption. For the sake of consistency, we use the same estimates of peripheral power consumption as \cite{6}. Crossbar power consumption in AIDX is computed as the average power consumed across the memristors in one inference operation. 
Area overhead is defined as the percentage increase in on-chip area required for the memristance drift solution due to peripherals, external circuit, and other items. Accuracy improvement is the increase in classification accuracy provided by a solution over the baseline model in a 1-hidden layer MLP at the end of the baseline model’s defined lifetime. Performance lifetime is defined as the amount of time required for a system to degrade to $70\%$ classification accuracy on the MNIST dataset. We chose this metric as an axis of comparison primarily because it is used in \cite{6} and is easily adaptable to the Interrupt and Benchmark method used in \cite{5}. 
Scalability is a measure of how well a memristance drift solution’s performance and overhead scales with crossbar size and additional layers in NN applications. Time overhead is not shown in Table 1 because there is negligible time overhead introduced by all solutions presented as compared to their respective baseline models. 


\begin{table}[!t]
	\centering
	\caption{Comparative analysis of AIDX.}
	\begin{tabular}{| >{\centering\arraybackslash}m{3.4cm} | >{\centering\arraybackslash \scriptsize}m{0.7cm}|>{\centering\arraybackslash \scriptsize}m{0.7cm}|>{\centering\arraybackslash \scriptsize}m{0.9cm}|>{\centering\arraybackslash \scriptsize}m{0.9cm}|} 
	\hline
	{\bf Methods} & \centering{\bf [5]} & \centering{\bf [6]} & \centering{\bf  AIDX-A} & {\bf AIDX-P} \\ 
	\hline 
	{\bf Power overhead (\%)} & - & $+1.61$ & $+3.27$ & $\mathbf{+1.19}$ \\ \hline  
	{\bf Area overhead (\%)}& $0$ & $+2.34$& $0$ & $0$ \\  
	\hline
	{\bf Performance life-time}& $1.22\times$ & $14.85\times$  & $\mathbf{37.62\times}$  & $31.41\times$  \\  
	\hline
	{\bf Scalability vs Baseline}& Worse & Worse &\bf Better &\bf Better\\  
	\hline
	{\bf Accuracy improvement (\%)}& $8.6$ & $37.3$ & $\mathbf{65.7}$ & $57.4$ \\ 
	\hline
	{\bf Include non-idealities}& No &  No &\bf Yes & \bf Yes \\  
	\hline
	\end{tabular}\label{tab: 1}
	\vspace{-6mm}
\end{table}
\section{Conclusion}
In this paper, we propose a new inference scheme based on voltage signal optimization called AIDX to reduce the impact of memristance drift on memristor crossbar MAC operations. By optimizing the voltage pulse width and amplitude input mapping, AIDX is flexible and effective across a different range of tasks including classification and image reconstruction. AIDX minimizes the computational error due to memristance drift. AIDX provides up to a $60\%$ and $78.60\%$ increase in classification accuracy on the CIFAR-10 datasets and image reconstruction of MNIST dataset, respectively. In addition, we have proposed an extension to the popular VTEAM model to more precisely simulate memristor behaviour below the switching voltage thresholds.
\vspace{-0.2cm}
\section*{Acknowledgment}
This work is supported by NSERC HIDATA project and ERC-CoG IONOS n773228.

\end{document}